\documentclass[xcolor=dvipsnames,a4,final,floatfix,amssymb,twocolumn,nofootinbib]{revtex4}

\usepackage[dvipsnames]{xcolor}
\usepackage{amstext}
\usepackage{amsmath}
\usepackage{latexsym}

\usepackage[english]{babel}
\usepackage[latin1]{inputenc}

\usepackage{times}
\usepackage[T1]{fontenc}

\usepackage{graphics}
\usepackage{graphicx}

\usepackage{setspace}

\usepackage{textpos}

\newcommand{\vs}{\vspace{1mm}}

\newcommand{\be}{\begin{equation}}
\newcommand{\ee}[1]{\label{#1} \end{equation}}
\newcommand{\ba}{\begin{eqnarray}}
\newcommand{\ea}[1]{\label{#1} \end{eqnarray}}
\newcommand{\nl}{\nonumber \\}

\newcommand{\pt}[2]{ \frac{{\textrm d} #1}{{\textrm d} #2}}
\newcommand{\dt}[2]{ {{\textrm d} #1}/{{\textrm d} #2}}

\newcommand{\df}[1]{ {\textrm d} { #1} }

\newcommand{\exv}[1]{{\, \left\langle {#1} \right\rangle \, }}
\newcommand{\xp}[1]{{\rm e}^{\, #1}}

\definecolor{Pergamen}{RGB}{235,225,200}
\definecolor{LightGray}{RGB}{235,235,230}
\definecolor{PaleBlue}{RGB}{190,210,255}
\definecolor{DarkGreen}{RGB}{0,80,20}
\definecolor{SoftRed}{RGB}{255,220,170}

\usepackage[english]{babel}
\usepackage[latin1]{inputenc}

\usepackage{times}
\usepackage[T1]{fontenc}

\usepackage{graphics}
\usepackage{graphicx}
\usepackage{fancybox}

\usepackage{setspace}

\usepackage{textpos}

\begin{document}
%%%%%%%%%%%%%%%%%%%%%%%%%% TITLE %%%%%%%%%%%%%%%%%%%%%%%%%%%%
\title[Reservoir Fluctuations] % (optional, use only with long paper titles)
{
%A Novel Entropy Formula based on %\\	% For what purpose?
%the Universal Thermostat Independence Principle
New Entropy Formula with Fluctuating Reservoir
}

%%%%%%%%%%%%%%%%%%%%%%%%%%% AUTHORS & INSTITUTES %%%%%%%%%%%%%%%%%%%%%%%%%%%
\author{T.S.~Bir\'o,  G.G.~Barnaf\"oldi and  P.~V\'an}
\affiliation{
  Heavy Ion Research Group\\
  MTA 
  %\hspace{2mm}
  %\raisebox{-3pt}{ \includegraphics[height=5mm]{wignerlogo.png} }
  %\hspace{2mm} 
  Wigner Research Centre for Physics, Budapest 
}

%%%%%%%%%%%%%%%%%%%%%%% DATES & COMMENTS %%%%%%%%%%%%%%%%%%%%%%%%%
\date{\today}

%%%%%%%%%%%%%%%%%%%%%%%%%%% AUTHORS & INSTITUTES %%%%%%%%%%%%%%%%%%%%%%%%%%%

\begin{abstract}
%\showthe\currentgrouplevel
%\showthe\currentgrouptype
Finite heat reservoir capacity, $C$, and temperature fluctuation, $\Delta T/T$, lead to 
modifications of the well known canonical exponential weight factor.
Requiring that the corrections least depend on the
one-particle energy, $\omega$, we derive a deformed entropy, $K(S)$.
%in the small one-particle energy expansion, $\omega \ll E$.
The resulting formula contains the Boltzmann\,--\,Gibbs, 
R\'enyi, and Tsallis formulas as particular cases.
For extreme large fluctuations, in the limit $C \Delta T^2 /T^2 \to \infty$, 
a new parameter-free entropy\,--\,probability relation is gained.
The corresponding canonical energy distribution is nearly Boltzmannian
for high probability, but
for low probability approaches the cumulative Gompertz distribution.
The latter is met in several phenomena, like earthquakes, demography, tumor growth models, 
extreme value probability, etc.

\pacs{05.70.-a,05.70.Ce,05.20.Gg}
\end{abstract}

%%%%%%%%%%%%%%%%%%%%%%%%%%% OUTLINE TOC %%%%%%%%%%%%%%%%%%%%%%%%%%%%
%\section*{}
%	\setcounter{tocdepth}{1}
%	\tableofcontents
%	\setcounter{tocdepth}{2}
  % You might wish to add the option [pausesections]

%%%%%%%%%%%%%%%%%%%%%%%%%%%%%%%%%%%%%%%%%%%%%%%%%%%%%%%%%%%%%%%%%%%%%%%%%%%%%%

%%%%%%%%%%%%%%%%%%%%%%%%%% GLOBAL OVERLAY SPECIFICATION %%%%%%%%%%%%%%%%%%%%%%%

% If you wish to uncover everything in a step-wise fashion, uncomment
% the following command: 

%\beamerdefaultoverlayspecification{<+->}

\maketitle

%%%%%%%%%%%%%%%%%% Preambulum slide (the Big Concept... ;-).  )
%{ \setbeamercolor{background canvas}{bg=Apricot}

%%%%%%%%%%%%%%%%%%%%%% INTRODUCTION %%%%%%%%%%%%%%%%%%%%%%%%%%%

%\section{Introduction}
\vs {\em Introduction} \vs

Presenting entropy formulas has a long tradition in
statistical physics and informatics.
The first, classical 'logarithmic' formula, designed
by Ludwig Boltzmann at the end of nineteenth century,
is the best known example, but -- often just out of mathematical
curiosity -- to date a multitude of entropy formulas are
known~\cite{KAZAN_BOOK,TANEJA}. Our purpose is not just
to add to this respectable list a number, we are after
some principles which would select out entropy formulas
for a possibly most effective incorporation of finite
reservoir effects in the canonical approach (usually
assuming infinitely large reservoirs). Naturally, this
endeavour can be done only approximately when restricting to
a finite number of parameters (setting $k_B=1$).

Among the suggestions going beyond the classical Boltzmann\,--\,Gibbs\,--\,Shannon
entropy formula,
\be
 S_B = - \sum_i p_i \ln p_i,
\ee{BGS_ENTROPY}
only a single parameter, $q$,  is contained in the R\'enyi formula~\cite{RENYI},
\be
S_R = \frac{1}{1-q} \, \ln \sum_i p_i^q.
\ee{RENYI_ENTROPY}
Many thoughts have been addressed to
the physical meaning and origin of the additional parameter, $q$,
in the past and recently.

The idea of a statistical -- thermodynamical origin
of power-law tailed distributions of the
one-particle energy $\omega$, out of a huge reservoir
with total energy, $E$ was expressed by using a power-law
form for the canonical statistical weight,
\be
w=\exp_q(-\omega/T) := \left(1 + (q-1) \frac{\omega}{T} \right)^{-\frac{1}{q-1}},
\ee{TS_WEIGHT}
instead of the classical exponential $\exp(-\omega/T)$\footnote{
The traditional exponential is restored in the $q\to 1$ limit.}.
Such weights can be derived from a canonical maximization
of the Tsallis-entropy~ \cite{TsallisOrigPaper,TsallisBook},
\be
 S_T = \frac{1}{1-q} \, \sum_i \left(p_i^q - p_i \right),
\ee{TS_ENTROPY}
or the R\'enyi-entropy eq.~(\ref{RENYI_ENTROPY}), too.
It is evident to justify that these two entropy formulas
are unique and strict monotonic functions of each other:
using the notation $C=1/(1-q)$, one easily obtains
\be
S_T = C \left( \xp{S_R/C}-1\right).
\ee{TS_RENYI}
The use of these entropy formulas is exact in case of
an ideal, energy-independent heat capacity reservoir~\cite{Almeida}.
The correspondence eq.~(\ref{TS_RENYI}) emerges naturally
from investigating  a subsystem \,--\, reservoir couple
of ideal gases~\cite{BiroPHYSA2013}.

Particle number or volume fluctuations in a reservoir
lead to further interpretation possibilities
of the parameter $q$~\cite{Wilk,Wilk2,Begun1,Begun2,Gorenstein,Gorenstein2}. 
In a recent paper~\cite{Biroarxiv2014} we demonstrated that both effects contribute
to the best chosen $q$ if we consider the power-law statistical
weight (\ref{TS_WEIGHT}) as a second order term in the
expansion in $\omega \ll E$ of the classical complement
phase-space formula, $w \propto \xp{S}$, due to Einstein.
A review of an ideal reservoir, with fixed energy, $E$,
and particle number, $n$, fluctuating according to the
negative binomial distribution (NBD), reveals that
the statistical power-law parameters are given by
$T=E/\langle n\rangle$ and 
$q=1 + \Delta n^2/\exv{n}^2 - 1/\exv{n}$.
The derivation relies on the evaluation of the microcanonical statistical factor, 
$(1-\omega/E)^n$, obtained as $\exp(S(E-\omega)-S(E))$, for ideal gases. 
Since each exponential factor grows like $x^n$, their ratio delivers the 
$(1-\omega/E)^n$ factor.
This factor is averaged over the assumed distribution of $n$.
The parameter $q$, obtained in this way is also named as second factorial moment, $F_2$, 
discussed with respect to canonical suppression in Refs.~\cite{KOCH,BEGUN}.
For the binomial distribution of $n$ one gets $q=1-1/k$, for
the negative binomial $q=1+1/(k+1)$.

The theoretical results on $q$ and $T$ depending on the mean
multiplicity, $\exv{n}$,  and its variance in the reservoir 
is just an approximation.
For non-ideal reservoirs described by a general equation of state,
$S(E)$, the parameter $q$ is given by
\be
q=1-1/C+\Delta T^2/T^2,
\ee{INTER_Q}
as it was derived in~\cite{Biroarxiv2014}.
It is important to realize that the scaled temperature variance
is meant as a variance of the fluctuating quantity $1/S'(E)$,
while the thermodynamical temperature is set by $1/T = \langle S'(E) \rangle$.
This effect and the finite heat capacity, $C$, act against each other.
Therefore even in the presence of these finite reservoir effects,
$q=1$ might be the subleading result, leading back to the use
of the canonical Boltzmann\,--\,Gibbs exponential.
In particular this is the case for the variance calculated in the
Gaussian approximation, when it is exactly $\Delta T/T = 1/\sqrt{|C|}$
and one arrives at $q=1$.
It is interesting to note that both parts of this formula,
namely $q=1-1/C$ and $q=1+\Delta T^2/T^2$, has been
derived and promoted in earlier 
publications~\cite{BiroPHYSA2013,Wilk4,Wilk5,Wilk6,BAGCI}.

In this paper we generalize the canonical procedure 
by using a deformed entropy $K(S)$~\cite{BiroPHYSA2013}. 
Postulating a statistical weight, $w_K$, based on $K(S)$ instead of $S$,
corresponding parameters, $T_K$ and $q_K$ occur.
We construct a specific $K(S)$ deformation function 
by demanding $q_K=1$. 
This demand can be derived from the requirement 
that the temperature set by the reservoir, $T_K$, is independent
of the one-particle energy, $\omega$.
We call this the {\em Universal Thermostat Independence}
Principle (UTI)~\cite{BiroEPJA2013}. 
The final entropy formula contains the Tsallis expression for $K(S)$ and the
R\'enyi one for $S$ as particular cases.
The Boltzmann--Gibbs formula is recovered at two special choices
of the parameters.
Surprisingly there is another limit, that of huge 
reservoir fluctuations, $C \Delta T^2/T^2 \rightarrow \infty$, 
when the low-probability tails, canonical to this
entropy formula, approach the cumulative Gompertz distribution,
$\exp(1-\xp{x})$~\cite{Gompertz,Casey,Lomnitz,Hirose}.

%%%%%%%%%%%%%%%%%%%%%%%%%%%%%%% SPECTRA vs MULTIPLICITY %%%%%%%%%%%%%%%%%

%%%%%%%%%%%%%%%%%%%%%%%%%%%%%%%%%%%%%%%%%%%%%%%%%%%%%%%%%

%\section{Fluctuations and Mutual Entropy}
\vs \vs {\em Fluctuations and Mutual Entropy} \vs
%%%%%%%%%%%%%%%%%%%%%%%%%%%%%%%%%%%%%%%%%%%%%%%%%%%%%%%%%%%%%%%%%%%%%%%%%%%

The description of thermodynamical fluctuations
is considered mostly in the Gaussian approximation. 
Reflecting the fundamental thermodynamic
variance relation, $\Delta E \cdot \Delta \beta = 1$ 
with $\beta=S'(E)$, the characteristic scaled
fluctuation of the temperature is derived~\cite{Uffink1,Lavenda,Uffink2}.
The variance of a well-peaked function
of a random variable is related to the variance of the original variable via the
Jacobi determinant, $\Delta f = |f'(a)| \Delta x$. Applying this to the
functions $E(T)$ and $\beta=1/T$, one obtains
$\Delta E = |C| \Delta T$ with the $C:=\dt{E}{T}$  definition of heat capacity,
and $\Delta \beta = \Delta T /T^2$. Combining these one obtains
the classical formula $\Delta T/T = 1/\sqrt{|C|}$.

%%%%%%%%%%%%%%%%%%%%%%%%%%%%%%%%%%%%%%%%%%%%%%%%%%%%%%%%%%%%%%%%%%%%%%%%

Traditionally statistical physics assumes that the
state space is uniformly populated considering a few
constraints on the totals of conserved quantities. But exactly
such constraints make expectation values and fluctuations
in the subsystem and in the reservoir statistically dependent.
Therefore not a product, but a convolution of phase space factors, $\rho$, 
describe such a couple of thermodynamical systems:
\be
\rho(E) = \int_0^{E}\limits\! \rho(E-\omega) \, \rho(\omega) \, \df{\omega}
\ee{OMEGA}
together with the form $\rho(E)=\xp{S(E)}$, leads to the normalized ratio
\be
1 = \int_0^E\limits\! \xp{S(E-\omega)+S(\omega)-S(E)} \, \df{\omega}.
\ee{RATIO_IS_ONE}
Viewing the integrand as a statistical weight factor, also used for
obtaining expectation values of $\omega$- or $E$-dependent
quantities of physical interest, one arrives at the interpretation
of the joint probability with the mutual entropy:
%\be
$ P = \xp{I(\omega; E)}$
%\ee{JOINT_PROB}
with
\be
I(\omega; E) = {S(\omega)+S(E-\omega)-S(E)} %=  \ln \frac{\rho(\omega) \rho(E-\omega)}{\rho(E)}.
\ee{MUTUAL_ENTROPY}
In the canonical situation the total energy $E$ is fixed and $\omega$ fluctuates;
so does the reservoir energy, $E-\omega$.
In the Gaussian approximation the mutual information factor, $I(\omega;E)$ is
evaluated in the saddle point approximation leading to the following general
property of the maximal probability state: From $I^{\prime}(\omega_*)=0$ one obtains
%\be
$S^{\prime}(\omega_*) =  S^{\prime}(E-\omega_*)$.
%\ee{ZEROTH}
Assuming small variance near this probability peak, the respective expectation
values of the derivatives, defined as the common thermodynamical temperature
in equilibrium, are also equal:
%\be
$1/T := \exv{S^{\prime}(\omega)} \approx S^{\prime}(\omega_*)$. 
%\ee{EQUIL}
The second derivatives, however, lead to an effective heat capacity
as the harmonic mean of the subsystem and reservoir heat capacities:
\be
 \frac{1}{C_*} := -T^2 I^{\prime\prime}(\omega_*) 
% = -T^2 S^{\prime\prime}(\omega_*) -T^2 S^{\prime\prime}(E-\omega_*) 
 = \frac{1}{C(\omega_*)} + \frac{1}{C(E-\omega_*)}.
\ee{HARMONIC}
This result is dominated by the smaller heat capacity, so there is
no use of expanding the one-particle phase space factor
$\rho(\omega)=\xp{S(\omega)}$. Only the rest can be safely expanded
with the canonical assumption, $\omega \ll E$:
\be
\xp{I} \approx \xp{S(\omega)} \, 
\left[ 1 - \omega S^{\prime}(E) + \frac{\omega^2}{2} \left[ S^{\prime}(E)^2 + S^{\prime\prime}(E) \right] \right]
\ee{CANO_PROB}
%%%%%%%%%%%%%%%%%%%%%%%%%
One possibility for going beyond the Gaussian approximation is
to investigate finite reservoir effects in the microcanonical
treatment~\cite{BAGCI,CAMPISI,UrmossyPLB2011,UrmossyPLB2012}. 
This is, however, usually quite entangled with
a complex microdynamical description of the interaction.
It is therefore of interest to find a beyond-Gaussian but canonical
approximation. 

Our idea is to construct such a $K(S)$ deformed entropy
expression, which compensates $q\ne 1$ effects in the $\omega \ll E$
expansion. In this way the probability weight factor of partitioning
the total energy $E$ to a sub-part $\omega$ and a rest of $E-\omega$,
%\be
$P \propto \xp{S(\omega)+S(E-\omega)-S(E)}$,
%\ee{PART}
is replaced by the more general form
\be
P_K \propto \xp{K(S(\omega))+K(S(E-\omega))-K(S(E))}.
\ee{KPART}
The one-particle phase-space factor, $\rho(\omega)\propto \xp{S(\omega)}$
is generalized to $\rho_K(\omega)\propto \xp{K(S(\omega))}$ in this formula.
The statistical weight factor is consisting of the rest: $w_K=P_K/\rho_K$.
Demanding now
\be
 \pt{^2}{\omega^2} \ln w_K = 0,
\ee{DEMAND}
we appeal to the Universal Thermostat Independence principle:
we wish to have the statistical weight for the one selected particle
with energy $\omega$ to be least dependent on the energy of that particle,
itself.  By annulating the second derivative % as in eq.~(\ref{DEMAND})
we reach this beyond the Gaussian level.

%%%%%%%%%%%%%%%%%% SUPERSTATISTICS FOR IDEAL GAS %%%%%%%%%%%%%%
We compare the traditional assumption, $K(S)=S$, and
the UTI principle, obtaining the optimal $K(S)$ to second order
in the canonical expansion.
We consider a general system with general reservoir fluctuations.
For small $\omega \ll E$
\ba
w&=&\exv{\xp{S(E-\omega)-S(E)}}_{\omega\ll E} = \exv{\xp{-\omega S^{\prime}(E)+\omega^2 S^{\prime\prime}(E)/2 - \ldots}}
\nl
&=& 1 - \omega \exv{S^{\prime}(E)} + \frac{\omega^2}{2} \exv{S^{\prime}(E)^2+S^{\prime\prime}(E)}+\ldots 
\ea{COMPLEMENT_PHASE_SPACE}
The power-law statistical weight (\ref{TS_WEIGHT}) to second order is
\be
w=\left(1+(q-1)\frac{\omega}{T} \right)^{-\frac{1}{q-1}} =
1-\frac{\omega}{T} + q \frac{\omega^2}{2T^2} - \ldots
\ee{TSALLIS_EXPAND_AGAIN}
Equating term by term, we interpret the statistical power-law parameters as
\be
\frac{1}{T} = \exv{S^{\prime}(E)} \quad {\rm and} \quad
q = \frac{\exv{S^{\prime}(E)^2 + S^{\prime\prime}(E)}}{\exv{S^{\prime}(E)}^2}.
\ee{INTERPRET}
A relation,  $\exv{S^{\prime\prime}(E)}=-1/CT^2$, 
follows from the definition of the heat capacity of the reservoir. 
%\be
%\frac{1}{C} = \pt{T}{E} = -T^2 \pt{}{E}\exv{S^{\prime}(E)} = -T^2 \exv{S^{\prime\prime}(E)}. 
%\ee{HEAT_SECONDER}
The UTI requirement eq.~(\ref{DEMAND}), when applied to the full form in 
eq.~(\ref{TSALLIS_EXPAND_AGAIN}), leads to $q=1$.
Summarizing, we acknowledge that the parameter $q$
has opposite sign contributions from $\exv{S^{\prime \: 2}}-\exv{S^{\prime}}^2$ 
and from $\exv{S^{\prime\prime}}$. In general $q$ is given by eq.~(\ref{INTER_Q})
up to second order. With this formula $q>1$ and $q<1$ are both possible.

%%%%%%%%%%%%%%%%%%%%%%%%%%%%%% SECTION IV %%%%%%%%%%%%%%%%%%%%%%%%%
%\vspace*{-3mm}
%\section{Deformed Entropy Formulas}
\vs \vs \vs {\em Deformed Entropy Formulas} \vs

Techniques to handle the $q=1$ case are known since long.
For dealing with $q \ne 1$ systems the calculations as a rule
are involved, but the introduction of 
a deformed entropy, $K(S)$, instead of $S$ provides
more flexibility for handling the subleading term in 
$\omega$~\cite{BiroEPJA2013,BiroPRE2011}. 
The deformed statistical weight has an average over the
reservoir fluctuations, as follows
\ba
w_K&=&\exv{\xp{K(S(E-\omega))-K(S(E))}}  \, = \, 1- \omega \pt{}{E} K(S(E)) 
\nl
\, &+& \, \frac{\omega^2}{2}
\left[	\pt{^2}{E^2} K(S(E)) + \left[\pt{}{E} K(S(E))\right]^2 \right]. 
\ea{DEFORMED_ENTROPY_STATISTICAL_WEIGHT}
Note that
$\pt{}{E} K(S(E)) = K^{\prime} S^{\prime}$ and
$ \pt{^2}{E^2} K(S(E)) = K^{\prime\prime} S^{\prime \, 2} + K^{\prime} S^{\prime\prime}$.
Comparing this expansion with the expression (\ref{TSALLIS_EXPAND_AGAIN}) we obtain the
parameters for the deformed entropy.
Using previous notations for averages over reservoir 
fluctuations but assuming that $K(S)$ is independent of these
we obtain
\ba
\frac{1}{T_K} &=& K^{\prime} \frac{1}{T},
\nl
\frac{q_K}{T_K^2} &=& \left(K^{\prime\prime}+K^{\prime \, 2} \right) \frac{1}{T^2} 
\left(1+\frac{\Delta T^2}{T^2} \right) - K^{\prime} \frac{1}{CT^2}.
\ea{KS_TASSILS_PAREMETERS}
By choosing a particular $K(S)$ one manipulates $q_K$.
%%%%%%%%%%%%%%%%%%%%%%%%%%%%%%%%%%%%%%%%%%%%%%%%%%%%%%%%%%%%%%%%%%%%%%%%%%%%%%%
After a simple division we obtain
\be
q_K = \left( 1 + \frac{\Delta T^2}{T^2} \right) 
\left( 1 + \frac{K^{\prime\prime}}{ K^{\prime \, 2}} \right) 
- \frac{1}{C} \frac{1}{K^{\prime}}.
\ee{qK}
Finally we gain a novel,  general deformed entropy formula including
the effect of reservoir fluctuations.
Demanding $q_K=1$, which is a simple consequence of eq.~(\ref{DEMAND}), 
one obtains the differential equation
\be
C \: \frac{\Delta T^2}{T^2} K^{\prime \, 2} -  K^{\prime}
+ C \: \left(1+\frac{\Delta T^2}{T^2} \right) K^{\prime\prime} = 0.
\ee{qK_ONE_DIFF_EQ}
The solution of eq.~(\ref{qK_ONE_DIFF_EQ}) to $K(0)=0, K^{\prime}(0)=1$ 
with $S$-independent $C$ and $\Delta T/T$ is given by
\be
K(S) = \frac{C_{\Delta}}{\lambda} \, \ln \left(1-\lambda + \lambda \xp{S/C_{\Delta}} \right).
\ee{K_FOR_qK_ONE}
Here $\lambda:=C\Delta T^2/T^2$ and $C_{\Delta}=C+\lambda$.
The composition rule for this quantity can be decomposed to two
simple steps:  defining $L(S)=C_{\Delta}\left(\xp{S/C_{\Delta}}-1\right)$,
the formal additivity, $K(S_{12})=K(S_1)+K(S_2)$, leads 
to\footnote{Here $S_1=S(E_1)$, $S_2=S(E_2)$, $S_{12}=S(E_1+E_2)$ and therefore
$S_{12}\ne S_1+S_2$ cf. eq.(\ref{MUTUAL_ENTROPY})}
\be
L(S_{12}) = L(S_1) + L(S_2) + \frac{\lambda}{C_{\Delta}} \: L(S_1) \cdot L(S_2).
\ee{L_TSALLIS_ADDI}
We point out that the non-additivity parameter in this formula is 
given by $\lambda/C_{\Delta}=\Delta T^2/(T^2+\Delta T^2)$,
for Gaussian scaling of the temperature fluctuations it is simply $1/(C+1)$.

Once having a $K(S)$ deformation function for the entropy,
one argues as follows. The $K(S)$ is constructed to lead to $q_K=1$ to
the best possible approximation. Therefore $K(S(E))$ is additive
for additive energy, $E$, to the same approximation.
Being additive, the addition can be repeated arbitrary times,
with a number $N_i$ of energies $E_i$ -- viewed as a statistical
ensemble. The occurence frequencies of a given energy $E_i$
are then well estimated by $p_i = N_i/N$ with $N=\sum_i N_i$
being the total number of occurences in the ensemble.
This quantity, $p_i$ is the usual approximation to the probability
of a state with energy $E_i$, hence one arrives at the 
construction formula~\cite{BiroPHYSA2013}
\be
 K(S) = \sum_i p_i \, K(-\ln p_i).
\ee{K_ADDITIVE}
Based on this, the following generalized entropy formula arises
for an ideal finite heat bath with fluctuations:
\be
K(S) = \frac{C_{\Delta}}{\lambda} \, \sum_i p_i \,
 \ln \left(1-\lambda + \lambda p_i^{-1/C_{\Delta}} \right).
\ee{GENERAL_TSALLIS_qK_ONE}

For $\lambda=C\Delta T^2/T^2=1$ the deformed entropy expression 
(\ref{GENERAL_TSALLIS_qK_ONE}) 
leads exactly to the Boltzmann entropy,
irrespective of the value of $C_{\Delta}$.
The same limit is achieved for infinite reservoirs, $C\to\infty$ while keeping
$\lambda$ finite; the entropy formula is traditional.

Not considering superstatistical, event-by-event fluctuations 
in the reservoir one assumes $\lambda=0$.
With such assumptions from $q_K=1$ we arrive at the original 
UTI equation~\cite{BiroEPJA2013}:
\be
\frac{K^{\prime\prime}}{K^{\prime}} = \frac{1}{C}.
\ee{UTI_EQUATION}
The solution of eq.~(\ref{UTI_EQUATION}) with $K(0)=0$ and $K^{\prime}(0)=1$ delivers
%\be
$K(S) = C \left(\xp{S/C}-1 \right)$
%\ee{UTI_SOLUTION}
and one obtains upon using $K(S)=\sum_i p_i K(-\ln p_i)$
the statistical entropy formulas of Tsallis and R\'enyi:
\be
K(S) = \frac{1}{1-q} \sum_i \left(p_i^{q}-p_i\right) \quad \textrm{and} \quad
S = \frac{1}{1-q} \ln \sum_i p_i^{q}.
\ee{RENYI_TSALLIS}

For huge fluctuations, $\lambda = C\Delta T^2/T^2 \gg C > 1$, 
eq.(\ref{qK_ONE_DIFF_EQ}) reduces to $K^{\prime\prime}=-K^{\prime \, 2}$
and leads to the parameter free formula,
\be
K(S) = \ln\left(1+S\right) \, = \, \sum_i p_i \ln \left( 1 - \ln p_i \right)
\ee{KS_qK_ONE_LAMBDA_INFTY}
even for arbitrary $C(S)$ dependence.
The canonical $p_i$ distribution to this is
obtained by maximizing $K(S)$
with the constraints $\sum_ip_i=1$ and $\sum_i p_i\omega_i = U$.
This Jaynes principle leads to
\be
 \pt{}{p_i} K(S) = \ln(1-\ln p_i) - \frac{1}{1-\ln p_i} = \alpha + \beta \omega_i,
\ee{K_CANO}
having the Lambert-W function, defined as the $W(x)$ satisfying $W\xp{W}=x$,
as part of the solution:
\be
p_i = \exp \left( 1-\frac{1}{W\left(\xp{-(\alpha+\beta\omega_i)}\right)} \right) 
\ee{KQKCANO}
For high probability, $p_i \approx 1$, the quantity $-\ln p_i$
is small. In this approximation the deformed entropy formula,
eq.~(\ref{GENERAL_TSALLIS_qK_ONE}), 
gives back the traditional Boltzmann\,--\,Gibbs\,--\,Shannon
entropy, and the canonical distribution becomes the familiar exponential.
For the opposite extreme, i.e. dealing with very low probability high-energy tails,
$W$ is small, and one obtains
\be
p_i \approx \xp{-\xp{\alpha+\beta\omega_i}}.
\ee{GOMPERTZ}
This result reminds to the complementary cumulative Gompertz distribution, 
originally discovered in demographic
models~\cite{Gompertz}, and later used as a tumor growth
model~\cite{Casey}. This distribution also occurs
in studies of extreme value distributions, showing deviations from
scaling in the occurence frequencies of large magnitude earthquakes~\cite{Lomnitz}
or on other seizmological phenomena\cite{Hirose}.

%%%%%%%%%%%%%%%%%%%%%% ACKNOWLEDGEMENT %%%%%%%%%%%%%%%%%%%%%%
%\vspace{-3mm}
{\bf Acknowledgement} \quad  
     This work was supported by Hungarian OTKA grants NK77816, 
     K81161, K104260, NK106119 and NIH TET\_12\_CN-1-2012-0016. Author GGB also thanks
     the J\'anos Bolyai Research Scholarship of the Hungarian Academy of
     Sciences.

%%%%%%%%%%%%%%%%%%%%%%%%% BIBLIOGRAPHY %%%%%%%%%%%%%%%%%%%%%%

\end{document}